\documentclass[aps,prl,superscriptaddress,twocolumn]{revtex4-1}
\usepackage{amsmath}
\usepackage{amssymb}
\usepackage{graphicx}
\usepackage[usenames,dvipsnames]{xcolor}
\usepackage{tikz}
\usepackage{pgffor}
\usepackage{verbatim}
\usepackage{bbm}
\usepackage{float}
\usepackage{mathrsfs}
\usepackage{bbm}
\usepackage{multibib}
\usepackage[colorlinks, breaklinks=true,linkcolor=red, citecolor=blue, linktocpage=true]{hyperref}

\renewcommand{\>}{\rangle}

\begin{document}

\title{Anomaly indicators for time-reversal symmetric topological orders}

\author{Chenjie Wang}
\affiliation{Perimeter Institute for Theoretical Physics, Waterloo, Ontario N2L 2Y5, Canada}

\author{Michael Levin}
\affiliation{James Franck Institute and Department of Physics, University of Chicago, Chicago, Illinois 60637, USA}
\date{\today}

\begin{abstract}
Some time-reversal symmetric topological orders are anomalous in that they cannot be realized in strictly two-dimensions without breaking time reversal symmetry; instead, they can only be realized on the surface of certain three-dimensional systems. We propose two quantities, which we call {\it anomaly indicators}, that can detect if a time-reversal symmetric topological order is anomalous in this sense. Both anomaly indicators are expressed in terms of the quantum dimensions, topological spins, and time-reversal properties of the anyons in the given topological order. The first indicator, $\eta_2$, applies to bosonic systems while the second indicator, $\eta_f$, applies to fermionic systems in the DIII class. We conjecture that $\eta_2$, together with a previously known indicator $\eta_1$, can detect the two known $\mathbb Z_2$ anomalies in the bosonic case, while $\eta_f$ can detect the $\mathbb Z_{16}$ anomaly in the fermionic case. 
\end{abstract}

\maketitle

A useful way to characterize two-dimensional (2D) gapped quantum many-body systems is in terms of the properties of their anyon excitations. For systems with global symmetries, one can study both topological and symmetry properties of anyons. These properties are said to describe the {\it symmetry-enriched} topological (SET) order in the many-body system\cite{wen12,kitaev06,essin13,mesaros13,barkeshli14,tarantino16}.


An interesting aspect of symmetry-enriched topological orders is that some of them are {\it anomalous} in the sense that they cannot be realized in strictly 2D lattice systems. Instead, these SETs can only be realized at the surface of certain 3D gapped systems that respect the same symmetry\cite{vishwanath13,wangc-science, metlitski15,wangc13b,chen14a,bonderson13,fidkowski13,metlitski14}. This raises an important question: how do we determine whether a given SET is anomalous? This question has largely been answered for SETs with purely unitary symmetries\cite{barkeshli14,chen14,kapustin14,cho14,wangj15,bbc} but it remains open, in general, for SETs with anti-unitary symmetries like time reversal invariance\cite{metlitski13,wangc13,qi15,hermele15,senthil15}.

In this work, we consider the anomaly-detection problem for the simplest class of time-reversal symmetric SETs --- namely those whose {\it only} symmetry is time-reversal invariance. We consider both bosonic and fermionic systems. In the bosonic case, there are two known types of time-reversal anomalies ($\mathsf T$ anomalies), both of which take values in $\mathbb Z_2$\cite{vishwanath13,chen13}. These two types of anomalies are exemplified by the ``three-fermion'' and ``eTmT'' SETs\cite{vishwanath13,burnell14,wangc13}, respectively. It is known that the first type of $\mathsf T$ anomaly can be detected by the following quantity:
\begin{equation}
\eta_1 = \frac{1}{D}\sum_{a\in \mathcal C} d_a^2 e^{i\theta_a} \label{eta1}
\end{equation}
Here $\mathcal C$ denotes the set of anyons in a given SET, $d_a$ is the quantum dimension of anyon $a$, $\theta_a$ is the topological spin of $a$, and $D = \sqrt{\sum_a d_a^2}$ is the total quantum dimension. The reason that $\eta_1$ qualifies as an anomaly detector or, in our terminology, an {\it anomaly indicator}, is that it has two special properties: it only takes the values $\pm 1$ and if $\eta_1=-1$, then the corresponding SET is anomalous.\footnote{To see why $\eta_1 = -1$ implies an anomaly, recall that for any strictly 2D system, $\eta_1= e^{i2\pi c_-/8}$ where $c_-$ is the chiral central charge of the edge modes living on the boundary\cite{kitaev06}. Since $c_-$ is odd under time reversal, any strictly 2D time-reversal symmetric system must have $c_-=0$ and hence $\eta_1 = 1$.}

The anomaly indicator $\eta_1$ is very useful, but unfortunately no analogous quantities have been found for the second kind of $\mathsf T$ anomaly for bosonic systems, nor for $\mathsf T$ anomalies in fermionic systems. The main purpose of this work is to propose two such anomaly indicators: (i) $\eta_2$, which detects the second type of $\mathsf T$ anomaly in bosonic systems, and (ii) $\eta_f$, which detects the $\mathbb Z_{16}$ $\mathsf T$ anomaly in fermion systems of DIII class\cite{ff1,ff2}. While we are not able to prove that $\eta_2$ and $\eta_f$ are anomaly indicators, we will provide several pieces of evidence to this effect.

{\it Second anomaly indicator for bosonic systems.}---We propose that the second $\mathsf T$ anomaly for bosonic topological orders can be detected by the following indicator:
\begin{equation}
 \eta_2   = \frac{1}{D}\sum_{a\in \mathcal C}d_a \mathcal T_a^2 e^{i\theta_a} \label{eta2}
\end{equation}
Like $\eta_1$, we conjecture that $\eta_2$ can only take the values $\pm 1$, and if $\eta_2 = -1$, then the SET is anomalous\footnote{Our conjecture is agnostic as to whether $\eta_1 = -1$ or $\eta_2 = -1$ is a {\it necessary} condition for being anomalous: in principle there could be additional types of $\mathsf T$ anomalies beyond $\eta_1$ and $\eta_2$, e.g. like the $H^3(G,A)$ type anomalies for unitary symmetries\cite{barkeshli14,chen14}.}.

In Eq.~(\ref{eta2}), we have introduced a new quantity, $\mathcal T^2_a$. Defining it requires two steps. First, recall that the time reversal operator $\mathcal T$ can {\it permute} different species of anyons. We denote this permutation by $a\rightarrow \mathcal T(a)$. Next, consider the subset of anyons satisfying $\mathcal T(a) = a$, i.e. the anyons that are invariant under the $\mathcal T$ permutation. Invariant anyons can be divided into two classes: those that carry a two-fold time-reversal protected Kramers degeneracy, similar to that of a spin-$1/2$ electron, and those that do not carry such a degeneracy.\footnote{For a precise definition of Kramers degeneracy in the context of anyons, see Ref.~\onlinecite{levin12b}.} We will say that an invariant anyon $a$ is a {\it Kramers doublet} if it belongs to the first class and a {\it Kramers singlet} otherwise. With this terminology, we define the quantity $\mathcal T^2_a$ as follows:
\begin{equation}
\mathcal T^2_a = \left\{
\begin{array}{ll}
1, & \ \text{if $\mathcal T(a) = a$, and Kramers singlet} \\[5pt]
-1, & \ \text{if $\mathcal T(a) = a$, and Kramers doublet} \\[5pt]
0, & \ \text{if $\mathcal T(a) \neq a$}
\end{array}
\right.\label{tsqure}
\end{equation}

When using Eq.~(\ref{eta2}), one should keep in mind that there are general physical constraints on
the $\mathcal T$ permutation and $\mathcal T^2$ assignments that hold for all SETs, whether or not they are
anomalous. The indicator $\eta_2$ is only applicable
when these constraints are satisfied. One example of such a constraint is that the topological spins must satisfy
$\theta_{\mathcal T(a)} = -\theta_a$ since $\mathcal T$ is anti-unitary.
Accordingly, all invariant anyons must have $\theta_a=0$
or $\pi$. Another constraint is that $\mathcal T$ cannot permute the trivial anyon $1$, i.e. $\mathcal T(1)=1$. In
addition, the trivial anyon must be a Kramers singlet, that is, $\mathcal T^2_1=1$. Likewise,
permuting an anyon twice should be trivial, so we have $\mathcal T[\mathcal T(a)]=a$. Lastly, in the
case of Abelian topological orders, both the $\mathcal T$ permutation and $\mathcal T^2$ assignments must respect
fusion rules in the sense that
\begin{align}
\mathcal T(a)\times\mathcal T(b) & = \mathcal T(a\times b), \quad \mathcal T_a^2 \mathcal T_b^2 = \mathcal T_{a\times b}^2 \label{prop}
\end{align}
where ``$\times$'' stands for the fusion product, and the second equation holds only for invariant anyons. Note that the above list is not exhaustive; for a more general discussion of constraints, see
Ref.~\onlinecite{barkeshli14}.

{\it Example.}---As an example, let us evaluate $\eta_2$ for the well-known toric-code topological order\cite{kitaev03}: $\mathcal C = \{1, e, m, \epsilon\}$. Here, $1$ is the trivial anyon, $e$ and $m$
are bosons, and $\epsilon$ is a fermion. All the anyons are Abelian, i.e., $d_a=1$ for every $a\in \mathcal C$. Accordingly,
the total quantum dimension is $D = 2$.  Consider the case that $\mathcal T$ does not permute anyons. Then, there are four
possible $\mathcal T^2$ assignments: $\mathcal T^2_e = \gamma_e$ and $\mathcal T^2_m = \gamma_m$, with
$\gamma_e,\gamma_m = \pm 1$ respectively. The trivial anyon must have $\mathcal T^2_1 =1$, and the fermion $\epsilon$ must
have $\mathcal T_\epsilon^2 = \gamma_e\gamma_m$. The latter follows from the fusion rule $e\times m =\epsilon$ and the
constraint (\ref{prop}). Inserting the above information into (\ref{eta2}), we obtain
\begin{equation}
\eta_2 = \frac{1}{2}(1 + \gamma_e + \gamma_m -\gamma_e\gamma_m)
\end{equation}
We observe that $\eta_2 =-1$ if $\gamma_e = \gamma_m =-1$ while $\eta_2 = 1$ otherwise. This agrees with expectations\cite{wangc13}:
the first case corresponds to the eTmT SET, which is believed to be anomalous, while the other three cases
are known to be non-anomalous, i.e. realizable in strictly 2D systems.

{\it Evidence.}---We now discuss the evidence for our conjecture about $\eta_2$.

{\it (1).} We have checked that $\eta_2=1$ for three large classes of strictly
2D systems:  (i) Kitaev's exactly soluble quantum double models with arbitrary finite group $G$ and with $\mathcal T$
acting like complex conjugation\cite{kitaev03}; (ii) double-layer topological orders $\mathcal B\times \bar {\mathcal B}$, where
$\mathcal B$ is an arbitrary bosonic topological order and $\bar{\mathcal B}$ is the time reversal partner of $\mathcal B$, and the two
layers are exchanged under $\mathcal T$ permutation;\footnote{We thank M. Metlitski for bringing this example into our attention.}
and (iii) Abelian topological orders described by $K$-matrix theory, discussed in Ref.~\onlinecite{levin12b}. We discuss details
of (i) and (ii) in the Supplementary Material\cite{sup}, and (iii) can be analyzed straightforwardly using the formula
(\ref{relation}) given below.

{\it (2).} We have checked that $\eta_2=-1$ for several systems that are believed to be anomalous. Examples that we considered include (i) the eTmT state discussed above, (ii) the (T-Pfaffian)$_-$ state, and (iii) four copies of the semion-fermion theory. While the latter two examples are fermionic systems --- in fact, they correspond to
SETs that live at the surface of 3D topological superconductors \cite{fidkowski13,metlitski14} --- they have bosonic counterparts
that can be constructed by gauging fermion parity symmetry. Our calculation is for these bosonic counterparts. We
present this calculation in the case of the (T-Pfaffian)$_-$ state in the Supplementary
Material\cite{sup}; the example (iii) can be treated in a similar fashion.

{\it (3).} We have checked that $\eta_2$ is {\it multiplicative} under stacking of topological orders. To see this, consider two bosonic topological
orders $\mathcal C$ and $\mathcal C'$, with total quantum dimensions $D$ and $D'$ respectively. In the stacked system
$\mathcal C \otimes \mathcal C'$, anyons are labeled by $(a,a')$ with $a\in \mathcal C$ and $a'\in \mathcal C'$. One can see that
$d_{(a,a')} = d_ad_{a'}$, $\theta_{(a,a')}= \theta_a+\theta_{a'}$, and the
total quantum dimension of $\mathcal C\otimes \mathcal C'$ is $DD'$. Also, $(a,a')$ is invariant under the $\mathcal T$ permutation if and only if both $a$ and
$a'$ are invariant, and $\mathcal T^2_{(a,a')} = \mathcal T^2_a \mathcal T^2_{a'}$. Putting this all together it
follows that $\eta_2$ (as well as $\eta_1$) is multiplicative under stacking. To see why this result is consistent with
expectations, recall that anomalous SETs are believed to be realized at the surface of 3D symmetry-protected topological (SPT) phases\cite{vishwanath13,chen13}.
Furthermore, it has been argued that 3D bosonic SPT phases with time reversal symmetry form a $\mathbb Z_2\times \mathbb Z_2$ group under
stacking. Therefore, we expect that the indicators $(\eta_1,\eta_2)$ should also form a $\mathbb Z_2\times\mathbb Z_2$ group under stacking.
In particular, $\eta_1,\eta_2$ should be multiplicative under stacking, as we just verified.

{\it (4).} In the case of Abelian topological orders, we have checked that $\eta_2$ does not change under a large class of topological
phase transitions, namely those arising from {\it anyon condensation}\cite{bais09} (see Supplementary Material\cite{sup}).
To understand why this property supports our conjecture, note that anomalies can be thought of as properties of {\it 3D bulk} phases
whose surfaces support anomalous SETs. On the other hand, topological phase transitions can be thought of as occurring on the {\it surface}.
Since surface phase transitions cannot change bulk properties, anomaly indicators must be invariant under such transitions.

{\it Alternative formula for $\eta_1\eta_2$.}--- In order to describe some additional evidence for our conjecture, we now discuss an alternative formula for the product indicator $\eta_1 \eta_2$. This formula is not as general as (\ref{eta1}) and (\ref{eta2}) and only applies to the case of Abelian topological orders. It states that $\eta_1 \eta_2$ can be computed as
\begin{equation}
\eta_1\eta_2 = e^{i\theta_a}\label{relation}
\end{equation}
where $a$ is any anyon that obeys
\begin{equation}
e^{i\theta_{a, b}} = \mathcal T_b^2 \quad \text{for all $b\in \mathcal I$} \label{eq-omega}
\end{equation}
Here $\mathcal I$ denotes the set of anyons that are invariant under the $\mathcal T$ permutation and $\theta_{a, b}$ denotes the mutual statistics between $a$ and $b$.

Before we derive Eq.~(\ref{relation}), let us discuss its implications. First, we can use it to show that $\eta_2$ can only take the values
$+1$ or $-1$: to see this, note that Eq.~(\ref{relation}) implies that $\eta_2$
has unit modulus. The claim then follows from the observation that $\eta_2$ is real.

Another interesting aspect of the formula (\ref{relation}) is that if we restrict to the case where $\mathcal T$
does not permute any anyons, then Eq.~(\ref{relation}) agrees with the more specialized time reversal anomaly formula conjectured in Ref.~\onlinecite{chen14}.\footnote{The latter formula was derived for anomalies associated with unitary symmetries, but conjectured to be true for anti-unitary symmetries as well.} This agreement provides further evidence for our conjecture.

We now turn to the justification of Eq.~(\ref{relation}).
We need to establish three points: {\bf (i)} there always exists at least one anyon $a$ satisfying Eq.~(\ref{eq-omega}); {\bf (ii)} if there exists multiple $a$'s satisfying Eq.~(\ref{eq-omega}), then they all share the same topological spin; and {\bf (iii)} the expression for $\eta_1 \eta_2$ in Eq.~(\ref{relation}) agrees with Eqs.~(\ref{eta1})-(\ref{eta2}). We prove the first two points in the Supplementary Material\cite{sup}. Here we will focus on the last point. To this end, we multiply Eqs.~(\ref{eta1}) and (\ref{eta2}) together and rewrite the resulting expression:
\begin{align}
\eta_1 \eta_2 & =  \frac{1}{D^2} \sum_c d_c^2 e^{i\theta_c} \sum_{b} d_b \mathcal{T}^2_b e^{-i \theta_b} \nonumber \\
&= \frac{1}{D^2} \sum_{abc}  e^{i\theta_c - i\theta_b} N^{{a}}_{{b} {c}} d_a d_c \mathcal{T}^2_b \nonumber \\
& = \frac{1}{D} \sum_{a} n_a d_a e^{i \theta_a}, \quad \quad n_a = \sum_b s_{ab} \mathcal{T}_b^2
\label{eta1eta2}
\end{align}
Here, the first equality follows from the fact that $\eta_2$ is real; the second equality follows from 
$d_b d_c = \sum_a N^{{a}}_{{b} {c}} d_a $; the third equality follows from the identity $
N^a_{bc} = N^{\bar{a}}_{\bar{b} \bar{c}} =  N^c_{a \bar{b}}$ together with the definition of the topological $S$-matrix\cite{kitaev06}: $s_{ab} = \frac{1}{D} \sum_{c} N^c_{a \bar{b}} e^{i\theta_c - i\theta_a - i \theta_b} d_c$.

So far, our computation of $\eta_1 \eta_2$ is completely general. If we specialize now to the Abelian case, then $s_{ab}$ reduces to $s_{ab} = e^{-i \theta_{a,b}}/D$, and we have the following identity:
\begin{equation}
n_a  = \left\{
\begin{array}{ll}
|\mathcal I|/D, & \ \text{if $a$ is a solution to (\ref{eq-omega})} \\[5pt]
0, & \ \text{otherwise}
\end{array}
\right.
\label{na}
\end{equation}
Here, to derive the `$0$' in the second line, we observe that both $\{\mathcal{T}_b^2\}_{b \in \mathcal I}$ and $\{e^{i \theta_{a,b}}\}_{b \in \mathcal I}$ define
one-dimensional representations of the subgroup $\mathcal I$ of invariant anyons. Therefore, by the orthogonality
properties of irreducible characters, $\sum_b e^{-i\theta_{a,b}} \mathcal{T}_b^2 = 0$ unless these
one-dimensional representations are equivalent, i.e. $a$ is a solution to (\ref{eq-omega}).

To complete the derivation, we substitute (\ref{na}) into (\ref{eta1eta2}). Using property {\bf (ii)} listed
above, we deduce that $\eta_1 \eta_2 = N|\mathcal I| e^{i\theta_a}/D^2$ where $a$ is any solution to (\ref{eq-omega}) and $N$
is the number of such solutions. At the same time, it is not hard to show that
$N = D^2/|\mathcal I|$. Eq.~(\ref{relation}) follows immediately.

{\it Anomaly indicator for fermionic systems.}---We now consider time-reversal symmetric topological orders in interacting fermionic systems of DIII class (i.e. those with $\mathcal T^2 = P_f$ where $\mathcal T$ and $P_f$ are the time-reversal and fermion parity operators). It is believed that the $\mathsf T$ anomalies in these systems have a $\mathbb Z_{16}$ structure under stacking, corresponding to the $\mathbb Z_{16}$ classification of 3D topological superconductors of DIII class.\cite{fidkowski13,metlitski14,wangc14,kitaev-z16} We propose that these $\mathsf T$ anomalies are detected by the following indicator:
\begin{equation}
\eta_f = \frac{1}{\sqrt 2 D} \sum_{a \in \mathcal C_f}d_a \tilde{\mathcal T}_a^2 e^{i\theta_a} \label{eta_f}
\end{equation}
We conjecture that $\eta_f$ can take $16$ different values, $e^{i\pi \nu /8}$ with $\nu =0, 1, \dots, 15$, and that the
SET is anomalous if $\eta_f \neq 1$.

Let us explain the expression (\ref{eta_f}). First of all, an essential difference between fermionic and bosonic topological orders is the existence of a {\it local} fermion $f$ in fermionic topological orders, which has trivial mutual statistics with all anyons and satisfies the fusion rule $f\times f=1$. We use $\mathcal C_f$ to denote the set of all anyons, including $f$. Anyons in $\mathcal C_f$ always come in pairs, \{$a$, $a\times f$\} where $a$ and $a \times f$ have topological spins that differ by $\pi$.

In Eq.~(\ref{eta_f}), we have introduced a new quantity $\tilde{\mathcal  T}_a^2$. To define it, we first introduce a related quantity which is given by
\begin{equation}
\mathcal T^2_a = \left\{
\begin{array}{ll}
1, & \ \text{if $\mathcal T(a) = a$, and Kramers singlet} \\[5pt]
-1, & \ \text{if $\mathcal T(a) = a$, and Kramers doublet} \\[5pt]
\pm i, & \ \text{ if $\mathcal T(a) = a\times f$} \\[5pt]
0, & \quad \text{ otherwise}
\end{array}
\right.
\label{T2fermion}
\end{equation}
(We will explain how to determine the signs in the $\pm i$'s below). With this definition,
$\tilde {\mathcal T}_a^2$ is given by:
\begin{equation}
\tilde{\mathcal T}_a^2 =\left\{
\begin{array}{cl}
-i\mathcal T^2_a, & \quad \text{ if $\mathcal T(a) = a\times f$} \\[5pt]
\mathcal T^2_a, & \quad \text{ otherwise}
\end{array}
\right. \label{ttilde}
\end{equation}
Here, the minus sign in the $-i$ in (\ref{ttilde}) is simply a matter of convention. In this convention, the surface of a DIII-class topological superconductor with index $\nu$ carries an anomaly $\eta_f = e^{i\nu\pi/8}$.
If instead we used $+i$ in (\ref{ttilde}), the indicator defined through (\ref{eta_f}) would be the complex conjugate of $\eta_f$ in the current convention.

We now explain how the $\pm i$'s in (\ref{T2fermion}) are assigned. This is subtle because when
$\mathcal T(a) = a\times f$, time reversal symmetry guarantees that $a$ and $a \times f$ are degenerate in energy.
Thus, $a$ and $a \times f$ {\it always} form a doublet. Nevertheless, previous work has shown that
the anyons obeying $\mathcal T(a) = a\times f$, can be divided into two classes which can be assigned the values
$\mathcal T^2_a = i$ and $\mathcal T^2_a = -i$ respectively.\cite{fidkowski13,metlitski14} Unlike Kramers doublets/singlets,
the physical distinction between anyons with $\mathcal T^2_a = \pm i$ is subtle, and the assignments depend on a sign convention; however, once a convention has been fixed, the $\mathcal T^2$ assignments are unambiguous.\footnote{For a precise definition of these $\mathcal T^2_a$ assignments, see Ref.~\onlinecite{metlitski14}.}

As in the bosonic case, the $\mathcal T$ permutation and $\mathcal T^2$ assignments must satisfy certain constraints. In particular, the relation $\theta_{\mathcal T(a)} = -\theta_a$ implies that the invariant anyons must have topological spin $\theta_a = 0$ or $\pi$ while the anyons with $\mathcal T(a) = a\times f$  must have $\theta_a = \pm \pi/2$. Also, the trivial anyon $1$ and the local fermion $f$ must be invariant under the $\mathcal T$ permutation, and must have $\mathcal T^2_1 =1$ and $\mathcal T^2_f =-1$. (The latter constraint follows from the definition of DIII-class fermionic systems).
Lastly, in the case of Abelian topological orders, there are constraints similar to Eq.~(\ref{prop}). However, instead of $\mathcal T^2_a$,  it is $\tilde {\mathcal T}^2_a$ that satisfies the relation $\tilde {\mathcal T}^2_a\tilde{\mathcal T}^2_b = \tilde{\mathcal T}^2_{a\times b}$, for all nonzero $\tilde{\mathcal T}^2_a$'s.\cite{fidkowski13,metlitski14}

{\it Examples.}---Let us evaluate $\eta_f$ for two examples. Our first example is the
so-called semion-fermion (SF) topological order. This system contains four Abelian anyons
$\{1, f, s, \bar s\}$, where $s$ is
a semion with $\theta_s = \pi/2$, and $\bar s = s\times f$ is an anti-semion with
$\theta_{\bar s} = -\pi/2$.  The $\mathcal T $ permutation takes $\mathcal T(s) = \bar s$
and $\mathcal T(\bar s) = s$.
As for the $\mathcal T^2$ assignments, we have $\mathcal T_f^2 =-1$, and $\mathcal T^2_1 = 1$ while
there are two possibilities for $\mathcal T^2_s$ and $\mathcal T^2_{\bar s}$, namely
$\mathcal T^2_s = -\mathcal T^2_{\bar s} = i\sigma$, with $\sigma=\pm 1$.
These two possibilities correspond to two types of semion-fermion topological orders known as
SF$_+$ and SF$_-$. Inserting
this information into (\ref{eta_f}) and using the definition (\ref{ttilde}) gives
\begin{align}
\eta_{f}\big|_{\rm SF_\sigma} = e^{i \sigma \pi/4}
\end{align}
This agrees with previous work which has argued that the SF$_+$ and SF$_-$ topological orders are
anomalous and live on the surfaces of $\nu=2$ and $\nu = 14$ topological
superconductors, respectively\cite{fidkowski13, metlitski14}.

Our second example is the $\rm SO(3)_6$ topological order\cite{fidkowski13}. This theory also contains
four anyons $\{1, f, s, \bar s\}$, with $\theta_s = \pi/2$ and $\theta_{\bar s} = -\pi/2$. The anyons $s$ and $\bar s$ are non-Abelian with $d_s = d_{\bar s} = 1+ \sqrt 2$. The $\mathcal T$
permutation is the same as in the semion-fermion topological order, and like that case
there are two variants of $\rm SO(3)_6$ with $\mathcal T_s^2 = -\mathcal T^2_{\bar s} = \pm i$. We will
refer to these two possibilities as $\rm SO(3)_{6,+}$ and $\rm SO(3)_{6,-}$. Substituting
this data into (\ref{eta_f}), we obtain
\begin{equation}
\eta_{f}\big|_{\rm SO(3)_{6,\sigma}} = e^{i\sigma 3\pi/8}
\end{equation}
where $\sigma = \pm 1$. Previous work has argued that the $\rm SO(3)_{6,\pm}$ topological orders are anomalous
and live on the surfaces of topological superconductors with odd index $\nu$, but the values of
$\nu$ have not been determined\cite{fidkowski13}. Our conjecture reveals these values: it implies that the $\rm SO(3)_{6,+}$
topological order lives on the surface of a $\nu=3$ topological superconductor, while
$\rm SO(3)_{6,-}$ lives on the surface of a $\nu=13$ topological superconductor.

{\it Evidence.}---We now turn to the evidence for our conjecture about $\eta_f$.

{\it (1).} We have checked that $\eta_f = 1$ for three large classes of strictly
2D fermionic topological orders. The first two classes are obtained by taking the 2D {\it bosonic} systems that we
discussed earlier --- namely (i) Kitaev's quantum double models and (ii) double layer bosonic topological
orders of the form $\mathcal B\times \bar {\mathcal B}$ --- and stacking them with a fermionic atomic insulator.
The third class consists of (iii) all fermionic Abelian topological
orders described by $K$-matrix theory\cite{levin12b}. Actually, the fact that $\eta_f = 1$ for
classes (i) and (ii) follows immediately from our previous result that $\eta_2 = 1$ for the corresponding bosonic systems, since
it is easy to show that $\eta_f = \eta_2$ for any fermionic system obtained by stacking a bosonic system with an atomic insulator.
As for class (iii), these systems can be analyzed via an alternative formula for $\eta_f$, similar to (\ref{relation}). This alternative formula is discussed in the Supplementary Material \cite{sup}.

{\it (2).} We have checked that $\eta_f \neq 1$ for several systems that are believed to be anomalous, including the (T-Pfaffian)$_-$ state, $N$ copies of the semion-fermion state ($N\notin 8\mathbb Z$), and $N'$ copies of the $\rm SO(3)_6$ state ($N'\notin 16\mathbb Z$)\cite{metlitski15,wangc13b,chen14a,bonderson13,fidkowski13,metlitski14}. On the other hand, we have checked that $\eta_f=1$ for the Moore-Read$\times U(1)_{-2}$ state, $T_{96}$ state, and (T-Pfaffian)$_+$ state from Refs.~\onlinecite{metlitski15,wangc13b,chen14a,bonderson13,fidkowski13,metlitski14}. This agrees with expectations since the latter topological orders are believed to be realizable in strictly 2D.

{\it (3).} We have checked that $\eta_f$ is {\it multiplicative} under stacking of topological orders.

{\it (4).} For the case of Abelian topological orders, we have checked that $\eta_f$ does not change under any topological phase
transition arising from anyon condensation (see Supplementary Material\cite{sup}).

{\it Discussion.}---To sum up, we propose two quantities, $\eta_2$ (\ref{eta2}) and $\eta_f$ (\ref{eta_f}),
for detecting anomalies in time-reversal symmetric bosonic SETs and DIII-class fermionic SETs, respectively.
Our proposal remains a conjecture. It is desirable to have a physical or mathematical proof. One possible
approach would be to construct, for each SET, a corresponding 3+1D topological field theory that supports
the SET on its 2+1D boundary. If the SET is {\it not} anomalous then the partition function of this 3+1D
theory should equal $1$ for every closed spacetime manifold. Thus, if one could show that the partition
function on some (non-orientable) closed manifold is equal to $\eta_2$ or $\eta_f$, then our
conjecture would follow\cite{witten16,metlitski15b}. Another possible approach would be to investigate
$\eta_2$ and $\eta_f$ in the context of 1+1D conformal field theory (CFT). Indeed, the relation
$\eta_1=e^{i2\pi c_-/8}$, which underlies the $\eta_1$ anomaly, was first proven in CFT\cite{kitaev06, frohlich90}. Hence, it seems plausible that relations analogous to $\eta_2 =1$ or $\eta_f = 1$ can also be derived in the context of time-reversal symmetric CFT.

{\it Acknowledgement.} We thank M. Cheng, M. Metlitski and Y. Qi for enlightening discussions. In particular, CW would like to
thank M. Metlitski for kindly teaching him various aspects of surface topological orders of 3D topological insulators/superconductors. ML is supported in part by NSF DMR-1254741. This research was supported in part by Perimeter Institute for Theoretical Physics. Research at Perimeter Institute is supported by the Government of Canada through the Department of Innovation, Science and Economic Development Canada and by the Province of Ontario through the Ministry of Research, Innovation and Science.

\bibliography{spt}

\clearpage

\onecolumngrid

\vspace{15pt}
\begin{center}
{\Large \bf Supplementary Material}
\end{center}
\vspace{30pt}

\setcounter{secnumdepth}{3}
\setcounter{equation}{0}
\renewcommand{\theequation}{\thesection\arabic{equation}}
\renewcommand{\thesection}{\Alph{section}}
\renewcommand{\thesubsection}{\arabic{subsection}}

\twocolumngrid

\section{Evaluation of $\eta_2$ in several examples}
\label{app1}

In this section, we evaluate $\eta_2$ for several examples.

\begin{table*}
\caption{List of $(e^{i\theta_a}, \mathcal T^2_a)$ for each anyon $a$ in the (T-Pfaffian)${_\pm}$ states (following Refs.~\onlinecite{fidkowski13,metlitski14}). Empty entries are not valid anyons. Here, $\gamma=+1$ corresponds to the (T-Pfaffian)$_+$ state and $\gamma=-1$ corresponds to the (T-Pfaffian)$_-$ state. } \label{tab1}
\centering
\begin{tabular}{|c|c|c|c|c|c|c|c|c|}
\hline
&  $\quad \ 0\quad\ $ & $\quad \ 1\quad \ $ & $\quad \ 2\quad \ $ & $\quad \ 3\quad \ $ & $\quad \ 4\quad\ $ & $\quad\ 5\quad \ $ & $\quad \ 6\quad \ $ & $\quad \ 7\quad \ $ \\
\hline
$I$  & $(+1,+1)$ &   & $ (-i, 0) $   &  & $(+1,-1)$  &  & $(-i,0)$ & \\
\hline
$\sigma$ &  & $(+1,+\gamma)$ &   & $ (-1,-\gamma)$   &  & $(-1,-\gamma)$  &  & $(+1,+\gamma)$ \\
\hline
$\psi$   & $(-1,+1)$ &   & $(+i, 0) $   &  & $(-1, -1)$  &  & $(+i, 0)$ & \\
\hline
\end{tabular}
\end{table*}

\subsection{Kitaev's quantum double models}

Kitaev's quantum double models are a class of 2D exactly soluble lattice models that support anyon excitations \cite{kitaev03}. The input for constructing a quantum double model is a finite group $G$, while the output is a generalized spin model where the spins live on the links of the square lattice and each spin can be in $|G|$ different states $|g\>$ where $g \in G$.

We only need a few properties of these models to compute $\eta_2$. First, the anyons can be labeled as pairs $a = ([g], \alpha)$ where $[g]$ is a conjugacy class of $G$ and $\alpha$ is an irreducible representation of the centralizer $Z_g=\{h|gh=hg, h\in G \}$. The quantum dimension and topological spin of the anyon $a$ are
\begin{equation}
d_a = |[g]| \cdot |\alpha|, \quad \quad e^{i\theta_a} = \frac{\chi_\alpha(g)}{|\alpha|}
\label{dgroup}
\end{equation}
where $|[g]|$ is the size of the conjugacy class, $|\alpha|$ is the dimension of the irreducible representation, and $\chi_\alpha$ is the character of the representation $\alpha$. Also, the total quantum dimension is known to be $D = \sqrt{\sum_a d_a^2} = |G|$.

Another property that we will need is that the quantum double models are {\it time-reversal symmetric} where time-reversal symmetry acts like ordinary complex conjugation in the $|g\>$ basis. The associated $\mathcal T$ permutation for the anyons is given by
\begin{equation}
([g], \alpha) \xrightarrow{\mathcal T} ([g], \alpha^*)
\end{equation}
where $\alpha^*$ is the complex conjugate representation of $\alpha$. Thus, an anyon $([g],\alpha)$ is invariant under the $\mathcal T$ permutation if and only if $\alpha$ is self-conjugate, i.e., $\alpha$ and $\alpha^*$ are equivalent irreducible representations of $Z_g$.

There are two kinds of self-conjugate representations, {\it real} and {\it pseudoreal} representations (the latter are also known as {\it quaternionic} representations). The $\mathcal T^2$ assignments for an anyon $a = ([g], \alpha)$ is determined by whether the corresponding representation $\alpha$ is real or pseudoreal:
\begin{equation}
\mathcal T^2_{a} = \nu_\alpha
\label{t2group}
\end{equation}
where
\begin{align}
\nu_\alpha &= \left\{ \begin{array}{ll}
1, & \text{ $\alpha$ real } \\
-1, & \text{ $\alpha$ pseudoreal } \\
0, & \text{ $\alpha$ not self-conjugate}
\end{array}
\right.
\end{align}
A few comments about Eq.~(\ref{t2group}): first, the quantity $\nu_\alpha$ is known as the {\it Frobenius-Schur} indicator. Second, while we will not include a derivation of Eq.~(\ref{t2group}), we should mention that this result can be obtained by applying the definition of Kramers degeneracy given in Ref.~\onlinecite{levin12b} to the anyon excitations of the quantum double model.

We are now ready to evaluate $\eta_2$. Plugging the expressions (\ref{dgroup}) and (\ref{t2group}) into (\ref{eta2}), and using the classical group theory result
\begin{align}
\nu_\alpha = \frac{1}{|H|}\sum_{h\in H} \chi_\alpha(h^2), \nonumber
\end{align}
we derive:
\begin{align}
\eta_2 & = \frac{1}{|G|} \sum_{[g],\alpha \in {\rm Rep}(Z_g)} |[g]| \cdot \nu_\alpha \cdot \chi_\alpha(g) \nonumber\\
& = \frac{1}{|G|} \sum_{g\in G} \sum_{\alpha \in {{\rm Rep}(Z_g)}} \frac{1}{|Z_g|}\sum_{h\in Z_g} \left(\chi_\alpha(h^2)\right)^* \chi_\alpha(g) \nonumber\\
& = \frac{1}{|G|} \sum_{g\in G} \sum_{h\in Z_g} \delta_{g,h^2} \nonumber\\
& =\frac{1}{|G|} \sum_{g\in G} \sum_{h\in G} \delta_{g,h^2} \nonumber\\
& =1
\end{align}
Here, to derive the second line, we use the above expression for $\nu_\alpha$ and we rewrite the summation over conjugacy classes $[g]$ as a summation over group elements $g$.
To get to the third line, we do the summation over $\alpha$ and use the orthogonality of characters. To get to the fourth line, we use the fact that all group elements $h\in G$ satisfying $h^2=g$ are contained in $Z_g$. The last line is straightforward if one does the summation over $g$ first.

We conclude that $\eta_2=1$ for Kitaev's quantum double models, in agreement with our conjecture.

\subsection{$\mathcal B\times\bar{\mathcal B}$ double-layer systems}
We can construct a large class of time-reversal symmetric 2D models by considering {\it double-layer systems} in which one layer is a spin model realizing some topological order $\mathcal B$ and the other layer is its complex conjugate in some local basis. To compute $\eta_2$ for such a system, note that its topological order is $\mathcal B\times\bar{\mathcal B}$ where $\bar{\mathcal B}$ is the time-reversal partner of $\mathcal B$. Thus, each anyon can be labeled as a doublet $(a,\bar b)$ where $(a,1)$ describes an excitation in the $\mathcal B$ layer and $(1, \bar b)$ describes an excitation in the $\bar {\mathcal B}$ layer. The quantum dimensions and topological spins of these anyons are
\begin{equation}
d_{(a, \bar b)} = d_a d_b, \quad e^{i\theta_{(a, \bar b)}} = e^{i\theta_a} e^{-i\theta_b},
\end{equation}
where $d_a$ and $e^{i\theta_a}$ are the quantum dimension and topological spin of $a \in \mathcal B$. The total quantum dimension is $D = \sum_{a\in \mathcal B} d_a^2$.

As for the time-reversal properties, one can check that these double-layer systems are invariant under a time-reversal operation which consists of complex conjugation followed by layer exchange. The associated time reversal permutation $\mathcal T$ exchanges the anyons $(a, \bar b)$ and $(b, \bar a)$, so the invariant anyons are of the form $(a,\bar a)$. Also, it is not hard to see that all the invariant anyons are Kramers singlets, i.e. $\mathcal T^2_{(a,\bar a)} =1$. Inserting all of this information into Eq.~(\ref{eta2}), we obtain
\begin{equation}
\eta_2 = \frac{1}{D}\sum_{a\in \mathcal B} d_a^2 = 1
\end{equation}
We conclude that $\eta_2=1$ for all double-layer states, which is again in agreement with our conjecture.

\subsection{Gauged T-Pfaffian states}

We start by reviewing the (T-Pfaffian)$_+$ and (T-Pfaffian)$_-$ states.
The (T-Pfaffian)$_+$ state is a fermionic topological order that can be realized in strictly 2D systems. The (T-Pfaffian)$_-$ state is a closely related {\it anomalous} fermionic topological order that can be realized at the surface of a $\nu=8$ 3D DIII topological superconductor\cite{metlitski15b}. Both states contain 12 anyons, including the local fermion. The anyons can be labeled as $I_k, \psi_k$ with $k=0,2,4,6$, and $\sigma_k$ with $k=1,3,5,7$ (Table \ref{tab1}). The anyons $I_k$ and $\psi_k$ are Abelian, while $\sigma_k$ are non-Abelian with quantum dimension $\sqrt 2$. The local fermion is $\psi_4$. The topological spins $e^{i\theta_a}$ and the values of $\mathcal T^2_a$ for both states are listed in Table \ref{tab1}, and under $\mathcal T$ permutation, $I_2\leftrightarrow\psi_2$ and $I_6\leftrightarrow\psi_6$. Note that the only difference between (T-Pfaffian)$_+$ and (T-Pfaffian)$_-$ is that they have different $\mathcal T^2$ assignments for the $\sigma$ anyons. In particular,
\begin{equation}
\mathcal T^2_{\sigma_1} = \mathcal T^2_{\sigma_7} = -\mathcal T^2_{\sigma_3} =-\mathcal T^2_{\sigma_5}=\gamma
\end{equation}
where $\gamma = \pm 1$ in the two cases.

Since the (T-Pfaffian)$_\pm$ states are built out of fermions\cite{chen14a,bonderson13}, they are not good examples for computing $\eta_2$. However, if we gauge fermion parity, the resulting gauge theories can be viewed as bosonic systems for which we can then calculate $\eta_2$ \cite{chen14a, fidkowski13}. (Note that it is important to gauge fermion parity in a time-reversal symmetric way, which is not always possible, but can be done for the (T-Pfaffian)$_\pm$ states.) Gauging fermion parity will introduce new anyons into the system, namely the {\it fermion parity fluxes}. The detailed properties of the gauged (T-Pfaffian)$_\pm$ states can be found in Ref.~\onlinecite{chen14a}, however, for our purpose we only need to know two facts about the gauged system: (1) after gauging, the total quantum dimension becomes $D= 4\sqrt 2 $, and (2) none of the flux excitations are invariant under $\mathcal T$ permutation. This second point is convenient for our calculation because it means that all nonvanishing $\mathcal T^2$ assignments are included in Table \ref{tab1}. Using these two facts and inserting Table \ref{tab1} into the expression for $\eta_2$ (\ref{eta2}), we obtain
\begin{equation}
\eta_2 = \frac{1}{D} 4\sqrt 2 \gamma =\gamma
\end{equation}
Thus, $\eta_2 = -1$ for the gauged (T-Pfaffian)$_-$ state and $\eta_2 = +1$ for the gauged (T-Pfaffian)$_+$ state. Again, this is consistent with our conjecture since previous work has argued that the gauged (T-Pfaffian)$_-$ state is anomalous while the gauged (T-Pfaffian)$_+$ state can be realized in strictly 2D systems\cite{metlitski15b}.


\setcounter{equation}{0}
\section{Solutions to Eq.~(\ref{eq-omega})}
\label{app2}

In this section, we discuss Eq.~(\ref{eq-omega}), which we reprint below for convenience:
\begin{equation}
e^{i\theta_{a,b}} = \mathcal T^2_b \quad \text{for all $b\in \mathcal I$} \label{s1}
\end{equation}
Here $\mathcal I$ denotes the set of invariant anyons in an Abelian topological order $\mathcal A$:
\begin{equation}
\mathcal I = \{b| \mathcal T(b) = b, b\in \mathcal A\}, \nonumber
\end{equation}
We will establish three points about the above equation: (1) solutions to (\ref{s1}) always exist, (2) given a solution $a$, every other solution $a'$ can be written as $a'= a\times x$, where  $x\in \mathcal E$ and $\mathcal E$ is the set
\begin{equation}
\mathcal E = \{ x | x = b\times\mathcal T (b), \ b\in \mathcal A \} \nonumber
\end{equation}
and (3) all solutions to (\ref{s1}) share the same topological spin.

To show the existence of solutions, we take a viewpoint from group representation theory. We note that the set of anyons $\mathcal A$ can be viewed as an Abelian group, with the group multiplication being the fusion product. The mutual statistics $\theta_{a,b}$ satisfies $e^{i\theta_{a,b}} e^{i\theta_{a,b'}} = e^{i\theta_{a,b\times b'}}$, so the set of phase factors $\{e^{i\theta_{a,b}}\}|_{b\in \mathcal A}$ can be viewed as a one dimensional representation of the group $\mathcal A$, with $a$ labeling the representation. In fact, by varying $a\in \mathcal A$, we go through {\it all} one dimensional representations of $\mathcal A$, and the correspondence between $a$ and one dimensional representations of $\mathcal A$ is one-to-one. This follows from the braiding non-degeneracy property of $\mathcal A$.

Next consider the subset of invariant anyons, $\mathcal I$. This subset is closed under fusion, hence it is a subgroup of $\mathcal A$. Therefore, the restriction $\{e^{i\theta_{a,b}}\}|_{b\in\mathcal I}$ defines a one dimensional representation of $\mathcal I$. By varying $a\in \mathcal A$, we again go through all one dimensional representations of $\mathcal I$, but the correspondence between $a$ and one dimensional representations of $\mathcal I$ is now many-to-one.

To complete the argument, note that the set $\{\mathcal T^2_b\}_{b\in\mathcal I}$ also defines a one dimensional representation of $\mathcal I$ since the $\mathcal T^2$ assignments obey the product rule $\mathcal T^2_b\mathcal T^2_{b'} = \mathcal T^2_{b\times b'} $. Hence there must exist at least one $a$ that corresponds to this one dimensional representation, i.e. at least one $a$ such that (\ref{s1}) is satisfied.

To show point (2), consider two solutions to (\ref{s1}), $a$ and $a'$. The fusion production $a'\times\bar a$ must have trivial statistics with all anyons in $\mathcal I$. Therefore, given a solution $a$ to (\ref{s1}), every other solution can be written as $a' =a\times x$, where $x\in \mathcal V$ and $\mathcal V$ is the subset
\begin{equation}
\mathcal V = \{x | \theta_{x,b} = 0, \text{ for every } b \in \mathcal I \} \nonumber
\end{equation}
Below we will show in a separate subsection (Appendix \ref{ev}) that $\mathcal V = \mathcal E$. Then, point (2) follows.

Finally, we show point (3), i.e, the topological spin factor $e^{i\theta_a}$ is the same for every $a$ satisfying Eq.~(\ref{eq-omega}). To see that, suppose $a'$ and $a$ are two solutions. Then, by point (2), we can write $a' = a \times x$,  where $x\in \mathcal E$. Accordingly, we have
\begin{align}
e^{i\theta_{a'}} &  = e^{i\theta_{a} + i\theta_x + i\theta_{a,x}}\nonumber\\
& = e^{i\theta_a} e^{i\theta_x} \mathcal T_x^2
\label{thetawprime}
\end{align}
where the first equality uses the relation $ \theta_{\alpha\times \beta} = \theta_\alpha+\theta_\beta+ \theta_{\alpha,\beta}$ for Abelian anyons, and in the second equality we have inserted Eq.~(\ref{s1}) and used the fact that
$\mathcal E \subset \mathcal I$. To proceed further, we use a nontrivial property of anyons $x \in \mathcal E$, namely that
\begin{equation}
\mathcal T_x^2 = e^{i\theta_x} \label{ep}
\end{equation}
That is, $\mathcal T^2_x$ is locked to the topological spin $e^{i\theta_x}$, for every $x \in \mathcal E$. This property is proved in Refs.~\onlinecite{bonderson13,chen14a,metlitski15,zaletel15,barkeshli14}. Substituting this relation into (\ref{thetawprime}), we immediately see that $e^{i\theta_{a'}} = e^{i\theta_a}$, which proves point (3).

\subsection{Showing $\mathcal E =\mathcal V$ }
\label{ev}

Above, we defined three subsets of $\mathcal A$, namely $\mathcal I$, $\mathcal E$ and $\mathcal V$. It is easy to check that all of them are actually subgroups of $\mathcal A$. Also, it is not hard to see that $\mathcal E \subset \mathcal I$ and $\mathcal E\subset\mathcal V$. Here we will show that $\mathcal E =\mathcal V$.

To do that, we first define the following auxiliary subsets:
\begin{align}
\mathcal I' & = \{b | \mathcal T(b) = \bar b, \ b\in \mathcal A\}\nonumber\\
\mathcal V' & = \{x | \theta_{x,b}=0, \ \text{ for every } b\in \mathcal I'\}\nonumber \\
\mathcal E' & = \{x | x = \bar b\times\mathcal T (b), \ b\in \mathcal A\}\nonumber
\end{align}
where $\bar{b}$ is the antiparticle of $b$. Again, $\mathcal I'$, $\mathcal V'$, and $\mathcal E'$ are subgroups of $\mathcal A$, with $\mathcal E' \subset \mathcal I'$ and $\mathcal E' \subset \mathcal V'$.

Next, we show the following relations
\begin{align}
|\mathcal V| & = |\mathcal A|/|\mathcal I| \label{c1} \\
|\mathcal V'| & = |\mathcal A|/|\mathcal I'| \label{c2}\\
|\mathcal E| & = |\mathcal A|/|\mathcal I'| \label{c3}\\
|\mathcal E'| & = |\mathcal A|/|\mathcal I| \label{c4}
\end{align}
where $|\dots|$ denotes the order of a set. To derive Eq.~(\ref{c1}), note that anyons in $\mathcal V$ are in one-to-one correspondence with one dimensional representations of $\mathcal A$ such that the elements in $\mathcal I$ are represented by the identity. The latter set of one dimensional representation are in turn in one-to-one correspondence with one dimensional representations of the quotient group $\mathcal A/\mathcal I$. Accordingly, $|\mathcal V| = |\mathcal A/\mathcal I| = |\mathcal A|/|\mathcal I|$. A similar argument leads to Eq.~(\ref{c2}). To show Eq.~(\ref{c3}), we define a map $\tau: \mathcal A \rightarrow \mathcal E$ by
\begin{equation}
\tau(b)= b\times \mathcal T(b)
\end{equation}
It is not hard to see that $\ker(\tau) = \mathcal I'$, and that $\tau$ is a surjective homomorphism. Therefore, we have $\mathcal A/\mathcal I' = \mathcal E$ which implies Eq.~(\ref{c3}). A similar argument leads to Eq.~(\ref{c4}).

With the help of the above relations (\ref{c1}) - (\ref{c4}), we can now show that $\mathcal E =\mathcal V$ using a counting argument. First, we note that $\mathcal E \subset \mathcal V$ so $|\mathcal E|\le |\mathcal V|$. According to Eqs.~(\ref{c1}) and (\ref{c3}), we then have
\begin{equation}
|\mathcal I|\le |\mathcal I'|
\end{equation}
Similarly, $\mathcal E'\subset \mathcal V'$, so $|\mathcal E'|\le |\mathcal V'|$. Then, according to Eqs.~(\ref{c2}) and (\ref{c4}), we have
\begin{equation}
|\mathcal I|\ge |\mathcal I'|
\end{equation}
Therefore, we have $|\mathcal I| = |\mathcal I'|$. Then, using (\ref{c1}) and (\ref{c2}), we have $|\mathcal E| = |\mathcal V|$ which implies that $\mathcal E=\mathcal V$.


\setcounter{equation}{0}
\section{Invariance of $\eta_2$ and $\eta_f$ under anyon condensation}
\label{app3}

In this appendix, we show that $\eta_2$ and $\eta_f$ are invariant under a large class of phase transitions --- namely those arising from {\it anyon condensation} in Abelian topological phases\cite{bais09}. This invariance under topological phase transitions is a necessary condition for $\eta_2$ and $\eta_f$ to be anomaly indicators, and thus provides additional evidence for our conjecture.

\subsection{Bosonic case}

To begin, let us briefly review the concept of anyon condensation. Anyon condensation is a kind of phase transition between topological orders. Consider an Abelian bosonic topological order with a set of anyons $\mathcal A$. Suppose that $\mathcal M\subset \mathcal A$ is a subgroup of anyons with the property that (1) all the anyons in $\mathcal M$ are bosons and (2) the mutual statistics between any two anyons in $\mathcal M$ is trivial. Then, we can imagine a scenario in which we tune some parameter in the Hamiltonian that decreases the mass gap for the anyons in $\mathcal M$, eventually causing the anyons in $\mathcal M$ to condense.

For our purposes, we are only interested in anyon condensation transitions that do not break time reversal symmetry. This puts two additional constraints on $\mathcal M$: (3) it should be closed under $\mathcal T$ permutation, i.e., if $m\in \mathcal M$, then $\mathcal T(m) \in \mathcal M$; (4) every invariant anyon $m \in\mathcal M$ must be a Kramers singlet, i.e., $\mathcal T_m^2 = 1$.

Two physical consequences of anyon condensation are as follows. First, certain anyons are {\it identified} after condensation. More specifically, if two anyons $a$ and $a'$ differ by some $m\in \mathcal M$, i.e., $a'=a\times m$, then $a$ and $a'$ will be identified as the same anyon after condensation. Second, if $a$ has nontrivial mutual statistics with some $m\in\mathcal M$, it will be {\it confined} after condensation.

Let $\mathcal L$ be the set of anyons that have trivial mutual statistics with all anyons in $\mathcal M$. Obviously, $\mathcal M$ is a subgroup of $\mathcal L$, which is in turn a subgroup of $\mathcal A$. Due to the identification and confinement phenomena mentioned above, the anyons in the condensed phase can be labeled by {\it cosets} $a\mathcal M$ where $a \in \mathcal L$.

What are the topological spins and time reversal properties of the anyons in the condensed phase? The topological spins are easy to compute: the topological spin of the anyon $a \mathcal M$ is given by $\theta_{a \mathcal M} = \theta_x$ for any $x\in a\mathcal M$. This is well-defined since one can check that all anyons $x\in a\mathcal M$ have the same topological spin --- assuming $\mathcal M$ obeys properties (1) and (2) listed above. The $\mathcal T$ permutation is also easy to compute since it is inherited from the corresponding permutation in the original phase. More specifically, in the coset labeling scheme, the $\mathcal T$ permutation is given by $\mathcal T(a\mathcal M) = \mathcal T(a) \mathcal M$. Here, we may choose any other representative $a'\in a\mathcal M$ and the $\mathcal T$ permutation is the same since $\mathcal T(a')\mathcal M = \mathcal T(a)\mathcal M$.

Determining the $\mathcal T^2$ assignments in the condensed phase is more subtle since in some cases these assignments are not uniquely fixed by the assignments before the condensation. To see how this works, note that a coset $a \mathcal M$ is invariant under the $\mathcal T$ permutation if for every $ x\in a \mathcal M$, there exists $m\in \mathcal M$ such that
\begin{equation}
\mathcal T(x) = x \times m,
\end{equation}
There are two cases to consider. The first case is that there is at least one $x\in a \mathcal M$ such that $\mathcal T(x) = x$. In this case, the $\mathcal T^2$ assignment is uniquely determined and is given by the corresponding assignment in the original phase: $\mathcal T^2_{a\mathcal M} = \mathcal T^2_x$. This assignment is well-defined since if there exists another anyon $x'\in a\mathcal M $ with $\mathcal T(x') = x'$, then one can check that $\mathcal T_{x'}^2=\mathcal T^2_x$. The second case is that $\mathcal T(x) \neq x$ for all $x \in a \mathcal M$. In this case, the value of $\mathcal T^2_{a\mathcal M}$ is not uniquely determined and may be either $\pm 1$; that is, there may be different types of condensation transitions that lead to different $\mathcal T^2$ assignments for $a \mathcal M$ \cite{burnell16}.

With this background, we are now ready to show that $\eta_2$ is invariant under every anyon condensation transition that preserves time-reversal symmetry. To do that, we first notice that $\eta_1$ is invariant under such transitions. This follows from the relation $\eta_1=e^{i2\pi c_-/8}$, where $c_-$ is the chiral central charge of the edge modes living on the boundary. It is known that $c_-$ does not change under anyon condensation, so $\eta_1$ does not change either. Accordingly, it suffices to show that the product $\eta_1\eta_2$ is invariant under anyon condensation transitions.

To establish the invariance of $\eta_1 \eta_2$, we need to show that $\eta_1 \eta_2 = \tilde\eta_1\tilde\eta_2$ where $\eta_1, \eta_2$ are the anomaly indicators in the original phase and $\tilde \eta_1, \tilde \eta_2$ are the indicators in the condensed phase. We do this using the formula (\ref{relation}) from the main text. Applying (\ref{relation}) to the uncondensed phase gives
\begin{equation}
\eta_1\eta_2 = e^{i\theta_a} \label{b1}
\end{equation}
where $a$ is any anyon that obeys
\begin{equation}
e^{i\theta_{a,b}} = \mathcal T^2_b, \quad \text{for all $b\in \mathcal I$} \label{b2}
\end{equation}
and where $\mathcal I$ is the set of invariant anyons. Similarly, applying the formula (\ref{relation}) to the condensed phase gives
\begin{equation*}
\tilde\eta_1\tilde\eta_2 = e^{i\theta_{a \mathcal M}}
\end{equation*}
where $a \mathcal M \subset \mathcal L$ is any coset that obeys
\begin{equation*}
e^{i\theta_{a \mathcal M, b \mathcal M}} = \mathcal T_{b\mathcal M}^2
\end{equation*}
for all invariant cosets $b \mathcal M \subset \mathcal L$. Equivalently, since $\theta_{a\mathcal M}=\theta_{a}$ and $\theta_{a \mathcal M,  b\mathcal M}=\theta_{a, b}$, we can rewrite the above formula for $\tilde\eta_1\tilde\eta_2$ as:
\begin{equation}
\tilde\eta_1\tilde\eta_2 = e^{i\theta_{a}} \label{b3}
\end{equation}
where $a$ is any anyon that obeys\footnote{Strictly speaking, we should require that $a \in \mathcal L$, but this requirement is implicit since $\mathcal T^2_{b\mathcal M}=1$ for $b\in\mathcal M$, so $a$ must belong to $\mathcal L$ as long as (\ref{b4}) is satisfied.}
\begin{equation}
e^{i\theta_{a, b}} = \mathcal T_{ b\mathcal M}^2, \quad \text{for all $b\in\tilde{\mathcal I}$} \label{b4}
\end{equation}
and where $\tilde{\mathcal I}$ is the set of anyons from all the invariant cosets $b\mathcal M \subset \mathcal L$.

Since the formulas (\ref{b1}) and (\ref{b3}) hold for {\it any} anyons $a$ obeying Eqs.~(\ref{b2}) and (\ref{b4}), we can prove $\eta_1\eta_2=\tilde\eta_1\tilde\eta_2$ if we can show that there is at least one anyon $a$ satisfying both (\ref{b2}) and (\ref{b4}). We now show that (\ref{b2}) and (\ref{b4}) indeed have common solutions. To do that, let us consider $\mathcal I\times \tilde{\mathcal I}$, the group of anyons obtained by fusing anyons from $\mathcal I$ and $\tilde{\mathcal I}$. Formally, we can assign a value of ``$T^2$'' to each anyon in $\mathcal I\times \tilde{\mathcal I}$, as follows: for $b\in \mathcal I$ and  $\tilde{b}\in \tilde{\mathcal I} $, we define
\begin{equation}
T^2_{b\times\tilde b} = \mathcal T^2_b \mathcal T^2_{\tilde b \mathcal M}
\end{equation}
Since $\mathcal T^2_x = \mathcal T^2_{x\mathcal M}$ for $x\in \mathcal I\cap \tilde{\mathcal I}$, these $T^2$ assignments are well defined. Now, suppose that we can find an anyon $a$
that obeys
\begin{equation}
e^{i\theta_{a,b}} =T^2_{b}, \quad \text{for all $b\in \mathcal I \times \tilde{\mathcal I}$} \label{b5}
\end{equation}
If such an anyon $a$ exists, then it automatically satisfies both (\ref{b2}) and (\ref{b4}). Indeed, this follows from the fact that $\mathcal I, \tilde{\mathcal I} \subset\mathcal I\times\tilde{\mathcal I}$ and $T^2_b = \mathcal T^2_b$ for $b\in \mathcal I$ and $T^2_{b} = \mathcal T^2_{b\mathcal M}$ for $b\in \tilde{\mathcal I}$. Thus, it suffices to show that there exists at least one anyon $a$ that obeys (\ref{b5}). This can be established using the same argument as in Sec.~\ref{app2} --- i.e. by observing that
$T^2_{b}$ defines a one dimensional representation of the group $\mathcal I \times \tilde{\mathcal I}$, and invoking the braiding non-degeneracy property of $\mathcal A$.

\subsection{Fermionic case}

We begin by reviewing how anyon condensation works in the fermionic case. Consider an Abelian fermionic topological order with a set of anyons $\mathcal A_f$. As in the bosonic case, we can condense any subgroup of anyons $\mathcal M \subset \mathcal A_f$ satisfying four properties: (1) every anyon in $\mathcal M$ is a boson, (2) the mutual statistics between any two anyons in $\mathcal M$ is trivial, (3) $\mathcal M$ is closed under $\mathcal T$ permutation, and (4) every invariant anyon $m\in {\mathcal M}$ obeys $\tilde{\mathcal T}^2_m =1$.

The properties of the anyons in the condensed phase are similar to the bosonic case. In particular, the anyons are labeled by cosets $a\mathcal M \subset \mathcal L$ where $\mathcal L$ is the set of anyons that have trivial mutual statistics with all anyons in $\mathcal M$. The topological spins of these anyons are given by $\theta_{a \mathcal M} = \theta_x$ for any $x\in a\mathcal M$. As for the $\mathcal T^2$ assignments, the rule is that $\tilde{\mathcal T}^2_{a\mathcal M} =\pm 1$ if the coset $a \mathcal M$ satisfies $\mathcal T(a \mathcal M) = a \mathcal M$ or $\mathcal T(a\mathcal M) = (a\times f)\mathcal M$. In the first case, the $\pm 1$ assignment is fully determined if $a\mathcal M$ contains at least one invariant anyon $x$, and is given by $\tilde{\mathcal T}_{a\mathcal M}^2 = \tilde{\mathcal T}^2_x$. In the second case, the $\pm 1$ assignment is fully determined if $a\mathcal M$ contains at least one anyon $x$ that satisfies $\mathcal T(x) = x\times f$ and is again given by $\tilde{\mathcal T}_{a\mathcal M}^2 = \tilde{\mathcal T}^2_x$.

With this preparation, we now show that $\eta_f = \tilde{\eta}_f$, where $\eta_f$ is the anomaly indicator in the original phase and $\tilde{\eta}_f$ is the indicator in the condensed phase. We will do this using two alternative formulas for $\eta_f$ and $\tilde{\eta}_f$, which we will justify below. In order to explain these formulas, we first need to recall a general fact about Abelian fermionic topological orders\cite{drinfeld09}: every Abelian fermionic topological order $\mathcal A_f$ can be written as a product $\mathcal A_f = \{1,f\}\times \mathcal A$, where $\mathcal A$ is a subgroup of anyons with {\it non-degenerate} braiding. Here the latter condition means that for every $a \in \mathcal A$ with $a \neq 1$ there exists some $b \in \mathcal A$ with $e^{i \theta_{a,b}} \neq 1$. The subgroup $\mathcal A$ is not necessarily unique, but we will assume a fixed choice for $\mathcal A$ in what follows.

With this notation, our formula for $\eta_f$ is
\begin{equation}
\eta_f = e^{i2\pi c_-/8} e^{-i\theta_a} \label{f1}
\end{equation}
where $a$ is any anyon in $\mathcal A$ that satisfies
\begin{equation}
e^{i\theta_{a, b}} = \tilde{\mathcal T}^2_b, \quad \text{for all $b\in \mathcal I$} \label{f2}
\end{equation}
Here $\mathcal I$ is defined by $\mathcal I =\mathcal I_f \cap \mathcal A$ where $\mathcal I_f$ is the set of anyons $b \in \mathcal A_f$ with
either $\mathcal T(b) = b$ or $\mathcal T(b) = b \times f$. Also, $c_-$ is the chiral central charge associated with $\mathcal A$, and $e^{i 2\pi c_-/8}$ is given by
\begin{equation}
e^{i2\pi c_-/8} = \frac{1}{\sqrt{|\mathcal A|}} \sum_{b\in \mathcal A} e^{i\theta_b} \label{f4}
\end{equation}
Our formula for $\tilde{\eta}_f$ is similar:
\begin{equation}
\tilde\eta_f = e^{i2\pi c_-/8} e^{-i\theta_{a}} \label{f5}
\end{equation}
where $a$ is any anyon in $\mathcal A$ that satisfies
\begin{equation}
e^{i\theta_{a,b}} = \tilde{\mathcal T}^2_{b{\mathcal M}}, \quad \text{for all $b\in \tilde{\mathcal I}$} \label{f6}
\end{equation}
Here $\tilde{\mathcal I}$ is defined by $\tilde{\mathcal I} = \tilde{\mathcal I}_f\cap \mathcal A$, where $\tilde{\mathcal  I}_f$ is the set of anyons from all cosets $a\mathcal M \subset \mathcal L$ with either $\mathcal T(a \mathcal M) = a \mathcal M$ or $\mathcal T(a \mathcal M) = f \times a \mathcal M$.

Before proving (\ref{f1}) and (\ref{f5}), let us see how they imply $\eta_f=\tilde\eta_f$. The argument is nearly identical to the bosonic case. The key point is that Eqs.~(\ref{f1}) and (\ref{f5})
hold for {\it any} anyons $a$ obeying (\ref{f2}) and (\ref{f6}), so we can prove $\eta_f=\tilde\eta_f$ if we can show that there exists at least one anyon $a$ that satisfies both (\ref{f2}) and (\ref{f6}). To prove that, we consider the group $\mathcal I \times \tilde{\mathcal I}$ and we formally assign ``$T^2$'' values to each anyon in $\mathcal I \times \tilde{\mathcal I}$ by defining
\begin{equation}
 T^2_{b\times\tilde b} = \tilde{\mathcal T}^2_b \tilde{\mathcal T}^2_{\tilde b\mathcal M}
\end{equation}
for each $b\in\mathcal I$ and $\tilde b\in \tilde{\mathcal I}$. Now suppose we can find an anyon $a \in \mathcal A$ that satisfies
\begin{equation}
e^{i\theta_{a,b}} =T^2_b, \quad \text{for all $b$'s in $\mathcal I \times \tilde{\mathcal I}$} \label{f8}
\end{equation}
If such an $a$ exists, then it is easy to see that it also satisfies (\ref{f2}) and (\ref{f6}). Therefore, all we have to do is show that (\ref{f8}) has at least one solution. This can be established straightforwardly using the fact that $T_b^2$ defines a one dimensional representation of the group $\mathcal I\times\tilde{\mathcal I}$ and the fact that $\mathcal A$ satisfies braiding non-degeneracy.

We now turn to the justification for (\ref{f1}) and (\ref{f5}). To prove (\ref{f1}), we need to establish three points: (i) solutions to (\ref{f2}) exist, (ii) all solutions to (\ref{f2}) share the same topological spin, and (iii) (\ref{f1}) agrees with Eq.~(\ref{eta_f}) from the main text. We follow similar arguments to those in Sec.~\ref{app2} to show these points. To show (i), we note that $\{e^{i\theta_{a,b}}\}|_{b\in \mathcal I}$ and $\{\tilde{\mathcal T}_b^2\}|_{b\in\mathcal I}$ define one-dimensional irreducible representations of the group $\mathcal I$ and also that anyons in $\mathcal A$ obey braiding non-degeneracy. To show (ii), we observe that given a solution $a$ to (\ref{f2}), any other solution $a'$ can be written as $a'=a\times x$, where $x$ is an anyon in $\mathcal A$ that has trivial statistics with respect to all anyons in $\mathcal I$. Furthermore, by following a similar counting argument as in Sec.~{\ref{ev}}, one can show that the set of anyons in $\mathcal A$ that have trivial statistics with respect to all anyons in $\mathcal I$ is the same as the set $\mathcal E = \mathcal E_f \cap \mathcal A$ where
\begin{equation}
\mathcal E_f  = \{x|x=\mathcal T(y) \times y, \text{ or } x =\mathcal T(y)\times y\times f, \ y\in \mathcal A_f \}
\label{ef}
\end{equation}
Finally, we observe that every anyon $x \in \mathcal E$ satisfies the relation $\tilde{\mathcal T}^2_x  e^{i\theta_x}=1$, which follows immediately from the property (\ref{ep}) and the fact that $\tilde{\mathcal T}^2_f=e^{i\theta_f}=-1$. Putting these three facts together, (ii) can be shown by performing a similar calculation as in (\ref{thetawprime}). Finally, to show (iii), we first rewrite the indicator $\eta_f$ in (\ref{eta_f}) as follows:
\begin{equation}
\eta_f= \frac{1}{\sqrt{|\mathcal A|}} \sum_{b\in \mathcal A} \tilde{\mathcal T}^2_{b}e^{i\theta_b} \label{f3}
\end{equation}
where we have used the fact that $D=\sqrt{2|\mathcal A|}$. Then, we multiply Eq.~(\ref{f4}) and the complex conjugate of Eq.~(\ref{f3}) and go through a similar calculation as in (\ref{eta1eta2}) and (\ref{na}). Straightforward algebra gives $\eta_f^* e^{i2\pi c_-/8} = e^{i\theta_a}$, where $a$ is any solution to (\ref{f2}). This establishes (\ref{f1}).

We now move on to proving (\ref{f5}). To do this, we again need to establish three points: (i) solutions to (\ref{f6}) exist, (ii) all solutions to (\ref{f6}) share the same topological spin,
and (iii) (\ref{f5}) agrees with the expression for $\tilde\eta_f$ obtained from applying Eq.~(\ref{eta_f}). Points (i) and (ii) can be argued similarly as above. To show point (iii), we rewrite
Eq.~(\ref{eta_f}) as
\begin{align}
\tilde\eta_f &= \frac{1}{\sqrt{2|{\mathcal L}|/|{\mathcal M}|}} \sum_{b{\mathcal M}\subset {\mathcal L}} \tilde{\mathcal T}^2_{b{\mathcal M}}e^{i\theta_b} \nonumber \\
 & = \frac{1}{|\mathcal M|\sqrt{2|{\mathcal L}|/|{\mathcal M}|}} \sum_{b\in{\mathcal L}} \tilde{\mathcal T}^2_{b{\mathcal M}}e^{i\theta_b} \nonumber\\
& = \frac{1}{\sqrt{|{\mathcal A}|}} \sum_{b\in \mathcal L \cap \mathcal A} \tilde{\mathcal T}^2_{b{\mathcal M}}e^{i\theta_b}  \label{f7}
\end{align}
Here, in the first line, we use the fact that $D=\sqrt{|\mathcal L|/|\mathcal M|}$ in the condensed phase. In the second line, we rewrite the summation over cosets $b\mathcal M \subset \mathcal L$ as a summation over all anyons in $\mathcal L$, and in the third line we use the fact that $|2\mathcal A| = |\mathcal A_f|=|\mathcal L||\mathcal M|$ and $\mathcal L = \{1,f\}\times (\mathcal L \cap \mathcal A)$. The last step is to multiply Eq.~(\ref{f4}) and the complex conjugate of Eq.~(\ref{f7}) and go through a similar calculation as in (\ref{eta1eta2}) and (\ref{na}). Straightforward algebra gives $\tilde\eta_f^* e^{i2\pi c_-/8} = e^{i\theta_{a}}$ where $a$ is any solution to (\ref{f6}). This proves (\ref{f5}).

\setcounter{equation}{0}
\section{Values of $\eta_f$ in Abelian topological orders}

In this section, we show that $\eta_f$ can only take values of the form $\{e^{i\nu \pi/4} | \nu \in \mathbb Z\}$ for Abelian topological orders.
This result is a corollary of the formula (\ref{f1}). To prove it, we note that if $a$ is a solution to (\ref{f2}), then the anyon $x=a\times a$ has trivial mutual statistics with all anyons $b$ in $\mathcal I$. This is because  $e^{i\theta_{x,b}} = [\tilde{\mathcal T}^2_b ]^2 = 1$ for all $b \in \mathcal I$. Now, as we mentioned in the previous section, any anyon $x$ that has trivial statistics with all anyons in $\mathcal I$ must belong to the set $\mathcal E = \mathcal E_f\cap \mathcal A$ where $\mathcal E_f$ is given in (\ref{ef}). Since $\mathcal E$ contains only bosons and fermions, we conclude that $e^{i\theta_x}=e^{i 4\theta_a} = \pm 1$. Therefore, we have
\begin{equation}
e^{i\theta_a} = e^{i m \pi/4}
\label{thetaa}
\end{equation}
where $m$ is an integer. Substituting (\ref{thetaa}) into (\ref{f1}) and using the fact that all Abelian topological orders have {\it integer} chiral central charge $c_-$, the result follows immediately.

\clearpage

\end{document}